\documentclass[a4paper]{jpconf}
\usepackage{graphicx}
\begin{document}
\title{Status of the SNO+ Experiment}

\author{E Caden\footnote{on behalf of the SNO+ Collaboration}}

\address{SNOLAB, 1039 Regional Road 24, Creighton Mine \#9, Lively ON P3Y 1N2, CANADA \\
		Laurentian University, 935 Ramsey Lake Road, Sudbury ON P3E 2C6, CANADA }

\ead{ecaden@snolab.ca}

\begin{abstract}
The SNO+ experiment is located at SNOLAB in Sudbury, Ontario, Canada. It will employ 780 tonnes of liquid scintillator loaded, in its initial phase, with 1.3 tonnes of $^{130}$Te (0.5\% by mass) for a low-background and high-isotope-mass search for neutrino-less double beta decay. SNO+ uses the acrylic vessel and PMT array of the SNO detector with several experimental upgrades and necessary adaptations to fill with liquid scintillator. The SNO+ technique can be scaled up with a future high loading Phase II, able to probe to the bottom of the inverted hierarchy parameter space for effective Majorana mass. Low backgrounds and a low energy threshold allow SNO+ to also have other physics topics in its program, including geo- and reactor neutrinos, supernova and solar neutrinos. This will describe the SNO+ approach for the double-beta decay program, the current status of the experiment and its sensitivity prospects.
\end{abstract}

\section{Introduction}
\begin{figure}[h]
\begin{minipage}{0.48\textwidth}
\includegraphics[width=\textwidth]{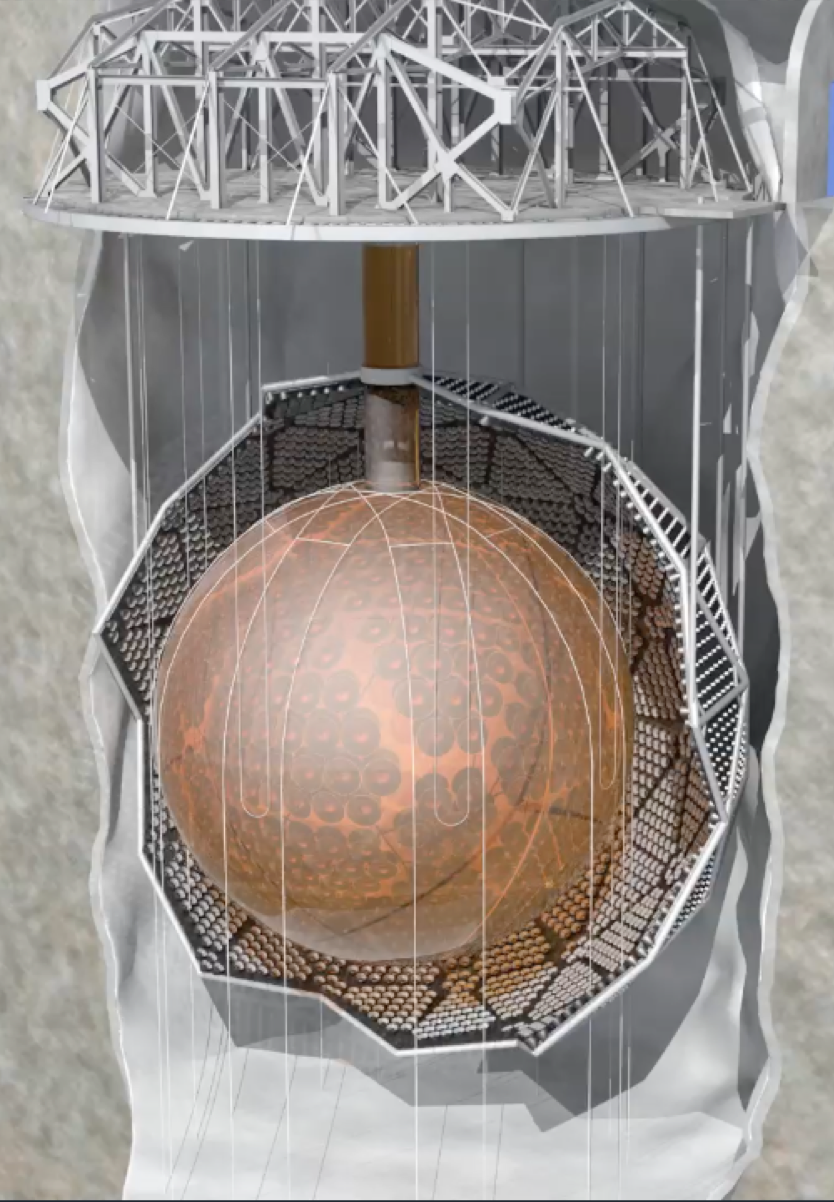}
\end{minipage}
\hspace{1pc}%
\begin{minipage}{0.48\textwidth}
\includegraphics[width=\textwidth]{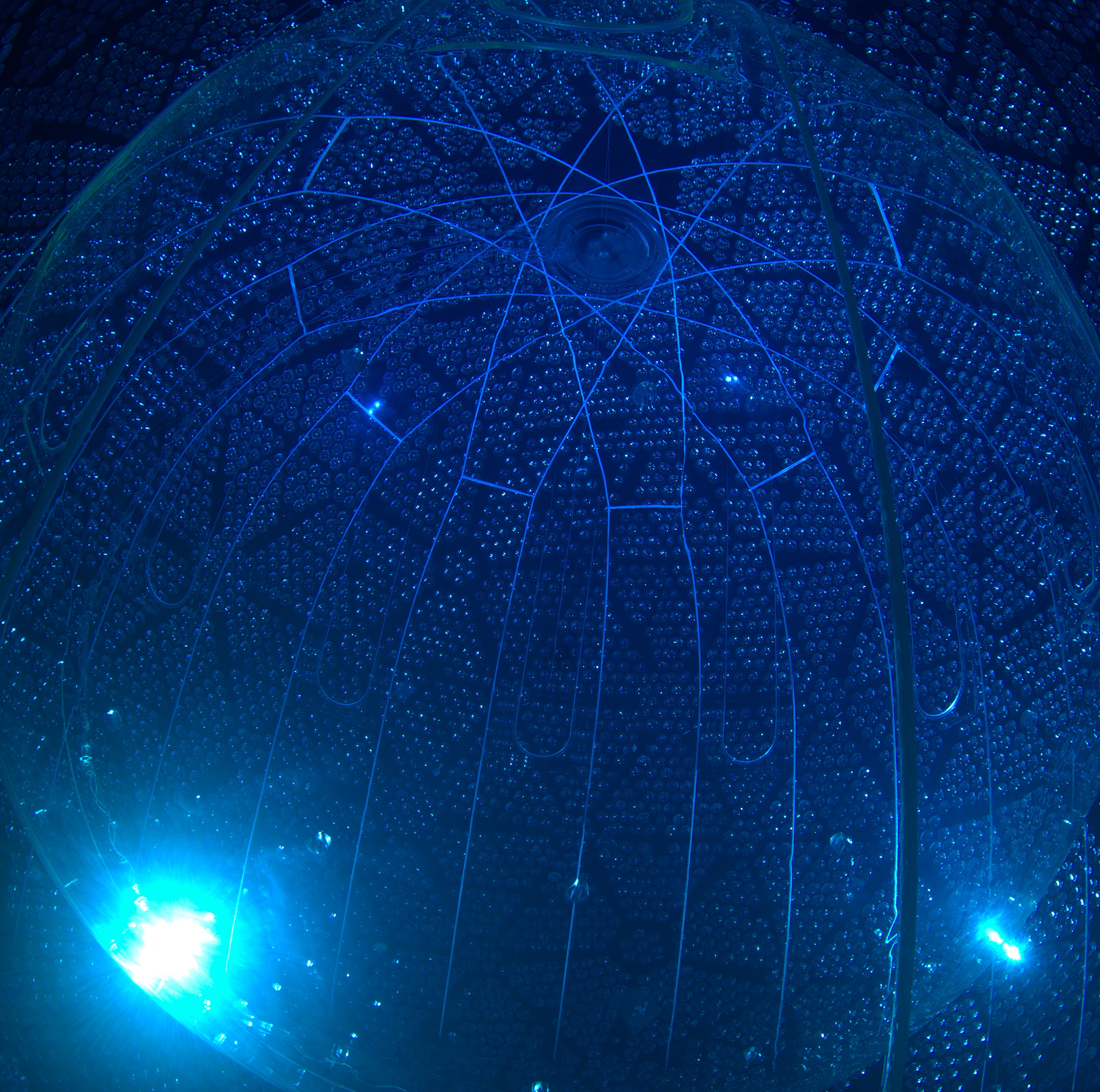}
\caption{\label{fig:detector} (left) An artist's depiction of the SNO+ detector in the original SNO cavern and (above) a photo of the detector. The new hold down net is visible, anchoring the detector to the floor when it is filled with liquid scintillator. }
\end{minipage}
\end{figure}

SNO+ is a double beta decay experiment currently operating in water phase at SNOLAB.  Its physics goals include detecting neutrinoless double beta decay using $^{130}$Te, solar neutrinos, reactor anti-neutrinos, geo anti-neutrinos, supernova neutrinos and performing exotic searches such as invisible nucleon decay \cite{review}. SNO+ reuses much of the existing infrastructure of SNO, consisting of a 12~m diameter acrylic vessel (AV) surrounded by almost 9,500 photomultiplier tubes (PMTs), shown in fig.~\ref{fig:detector}. A depth of 6000 m.w.e provides shielding from cosmic muons. 92 of those PMTs are outward facing and used to veto muons that interact in the 5300tonnes of external water shielding. There is an additional 1700 tonnes of water shielding between the AV and PSUP. In SNO+, the acrylic vessel will be filled with a liquid scintillator of linear alkyl benzene (LAB) and the fluor 2,5-diphenyloxazole (PPO) at 2~g/L, surrounded with a shielding of ultra-pure water. 

This includes construction of a process plant to distill and mix the scintillator cocktail before filling the detector \cite{scintplant}. The plant has a multi-stage distillation process that removes heavy metals and improves the UV transparency of the scintillator. The PPO is purified in a parallel distillation stream. Nitrogen steam stripping removes Rn, Kr, Ar and O$_2$, while water extraction removes Ra, K and Bi. Six parallel metal scavenger columns remove Bi, Pb, Ra, Ac, and Th. When the detector is operational, scintillator can be recirculated at 150~litres/minute to provide re-purification as necessary. Commissioning of the plant is ongoing, and filling with scintillator will happen immediately upon its completion.

\section{Detector Upgrades}
Significant work has taken place to transform the heavy water detector of SNO into a liquid scintillator detector. Scintillation reactions produce much more light than cherenkov reactions. For SNO+, this meant upgrading the trigger and data acquisition systems to handle the higher rates. New crate readout cards handle the higher data rates while and provide ethernet controls for the front-end electronic boards. Upgraded analog trigger cards have the ability to handle the larger current load created by the greater number of PMT hits. These also provide baseline stabilization and monitoring of triggers as well as the ability to introduce reprogrammable trigger logic. The CAEN v1720 digitizer adds the triggered waveforms to the event data, which will be used in the tagging of instrumental backgrounds. TUBii is an additional trigger utility board that provides many useful tools including pulsers and delays for calibration sources, extra trigger ports, a backup clock and dynamically reprogrammable logic of triggers.

The data acquisition system for SNO+ uses a modular approach that decouples the data flow from the detector control and monitoring tools, providing
greater stability and increased control. This has been running since the beginning of 2017. Graphical monitoring tools have been developed allowing realtime time-series
monitoring of data up to an event rate of 20kHz, which will be necessary during scintillator running. A slow control system provides web-based monitoring and alarms. The detector state for each run is recorded in a database, allowing for clear reproducibility of individual runs. The initial run of water phase has exercised the data flow and data processing systems. 

The SNO+ hold-down rope net was installed in 2012. This is required to counteract the buoyancy effect of the scintillator ($\rho=0.86$) in the acrylic vessel. This can be seen in figure~\ref{fig:detector}, around the next of the detector and anchored to the floor of the cavity. In 2016, tests were carried out to ensure this system responded correctly by mimicking the density change that will come when filling the AV with scintillator. The ropes behaved exactly as expected under the expected full upward force \cite{ropenet}.

Commissioning of the detector was completed in April 2017 and the water-fill phase of the experiment commenced on 4 May 2017.

\section{Calibration}
Calibration of the detector is crucial to understanding the performance of out detector. Specifically, we are measuring our energy scale and resolution, out position reconstruction and resolution, and PMT efficiency. Calibrations during water phase will feed into our understanding during scintillator phase, and those done during scintillator phase will contribute to understanding the detector during Te-loaded scintillator phase. SNO+ has both deployed and in-situ fibre calibration systems.

\subsection{Deployed Sources}
During water phase, calibration sources are deployed into the AV and inner water shielding areas using the original SNO source deployment hardware. These three Umbilical Retieval Mechanisms (URMs) were deconstructed and all their parts were cleaned and refurbished SNO+. In May 2017 the Laserball was first deployed in the center of the AV. the Laserball is a light diffuser used for studying detector's optical properties and PMT angular response, as well as letting us study our trigger efficiency. Its anisotropies were well documented in SNO. A new Laserball source with reduced shadowing and improved light uniformity will be used in the scintillator phase of SNO+.

In June 2017 a $^{16}$N source was first deployed along the vertical axis. This source gives calibration point at 5~MeV, which lets us measure our energy scale and our energy resolution.

SNO+ is also developing AmBe and Cherekov sources for deployment in both water and scintillator phases. Scandium sources will be used in the scintillator phase. New fully sealed source deployment hardware for the scintillator phases has been constructed and is currently being commissioned.

\begin{figure}[h]
\begin{minipage}{0.47\textwidth}
\includegraphics[width=\textwidth]{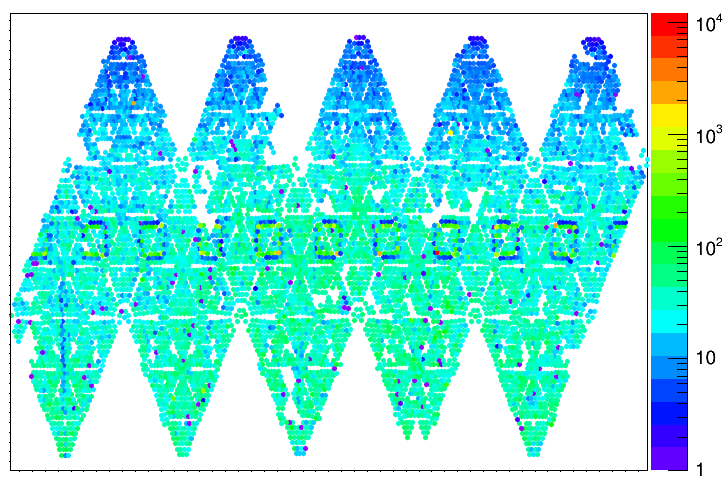}
\caption{\label{fig:laserball}A flat-map display of the Laserball run in the detector. The shadows of the fill-pipes and the hold-up ropes can be seen, as well as the self shadowing of the source.}
\end{minipage}\hspace{2pc}%
\begin{minipage}{0.52\textwidth}
\includegraphics[width=\textwidth]{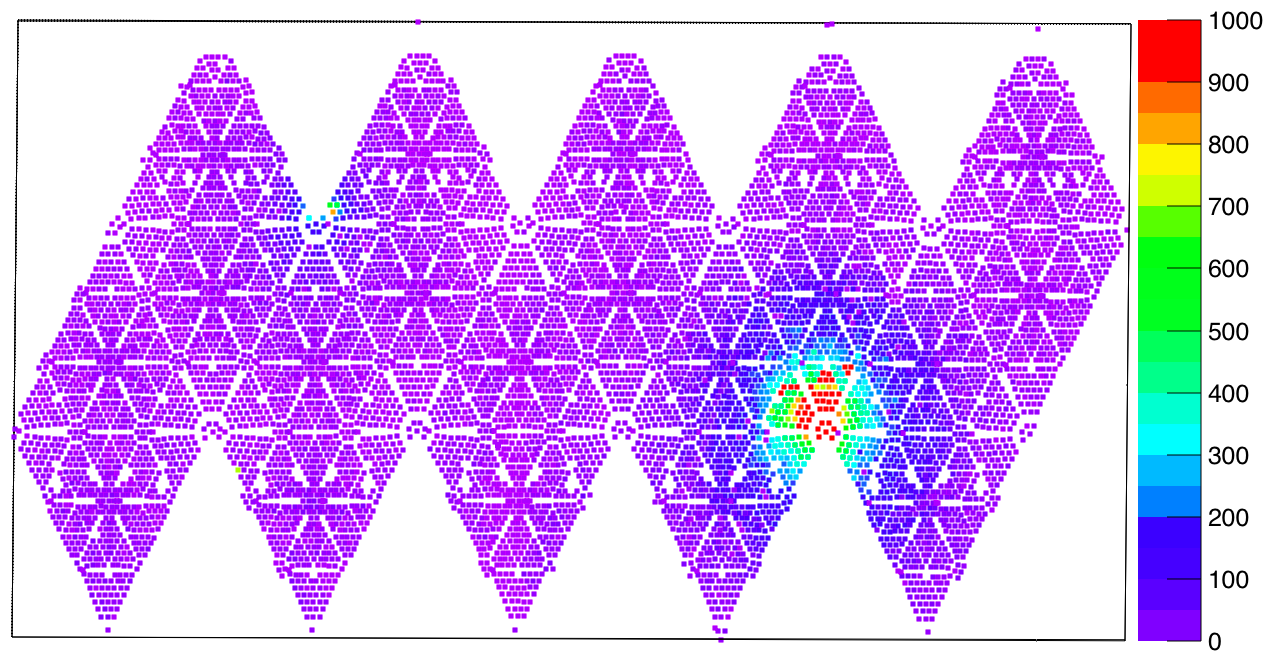}
\caption{\label{fig:smellie}A flat-map display of the SMELLIE calibration system, firing into the detector. The small beam spot is the light reflected by the AV, and the larger spot is directly across form the light injection point.}
\end{minipage} 
\end{figure}

\subsection{In-Situ Fibres}
SNO+ uses an in-situ fibre system to inject light from the PMT array to the detector using optical fibres. Each subsystem measures a different property of the inner detector medium: timing, scattering, and attenuation. Commissioning these systems in the water phase allows for better understanding of the detector before introducing the scintillator. These systems will allow for constant monitoring of the scintillator quality during Te-loaded phase, without possibly contaminating the detector. Commissioning of all three subsystems is ongoing.

\section{Te-Loading}
 The initial phase of SNO+ will have 0.5\% Te-loading by weight, which gives $\sim$ 1300 kg of $^{130}$Te. The telluric acid (TeA) for SNO+ has been underground ``cooling'' from cosmogenically-activated radiation since 2015. It will be further purified by multi-pass recrystallization, which is based on solubility of TeA in both water and nitric acid. Specifically, this removes contamination from $^{60}$Co, $^{110m}$Ag, $^{88}$Y and $^{22}$Na, which have long enough half-lives and whose decay energy overlaps the ($0\nu\beta\beta$) energy region-of-interest (ROI). This purification method was developed \cite{tepurification}, has been shown to be very effective in removing $^{60}$Co, with efficiencies of $10^6$ after two passes.

\begin{figure}[h]
\begin{minipage}{0.2\textwidth}
\includegraphics[width=\textwidth]{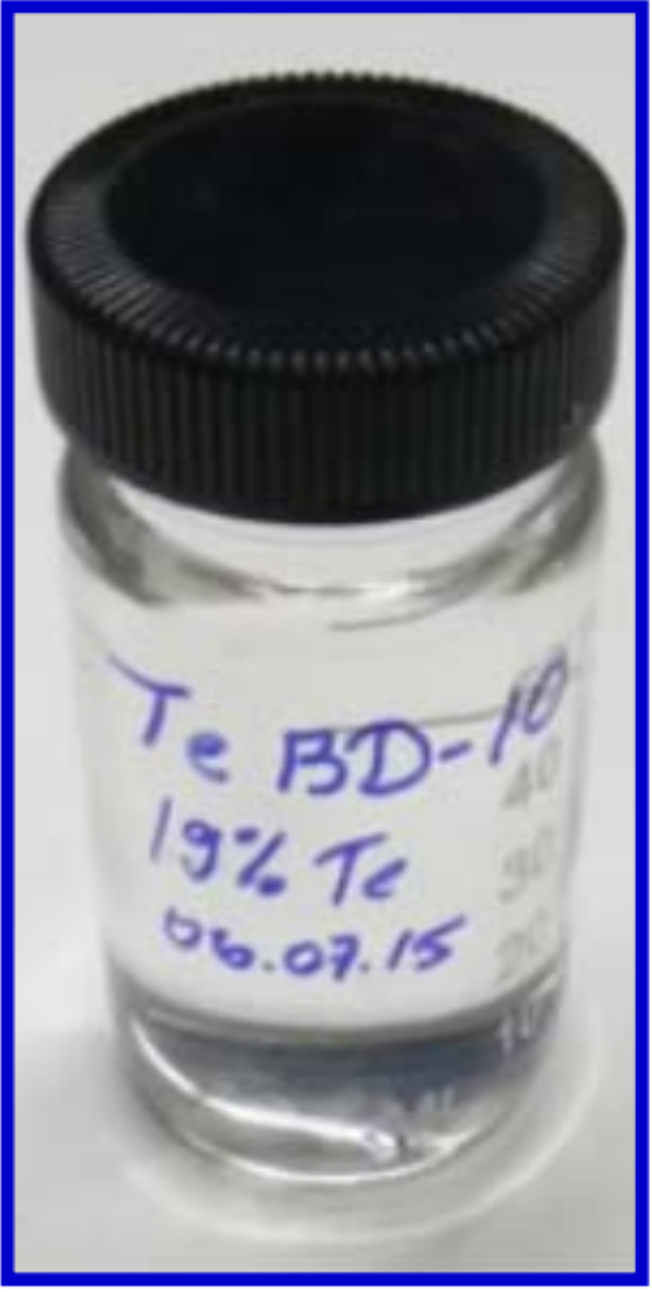}
\caption{\label{fig:TeBD}Vial of TeBD, showing high loading potential and optical clarity.}
\end{minipage}\hspace{3pc}%
\begin{minipage}{0.7\textwidth}
\includegraphics[width=\textwidth]{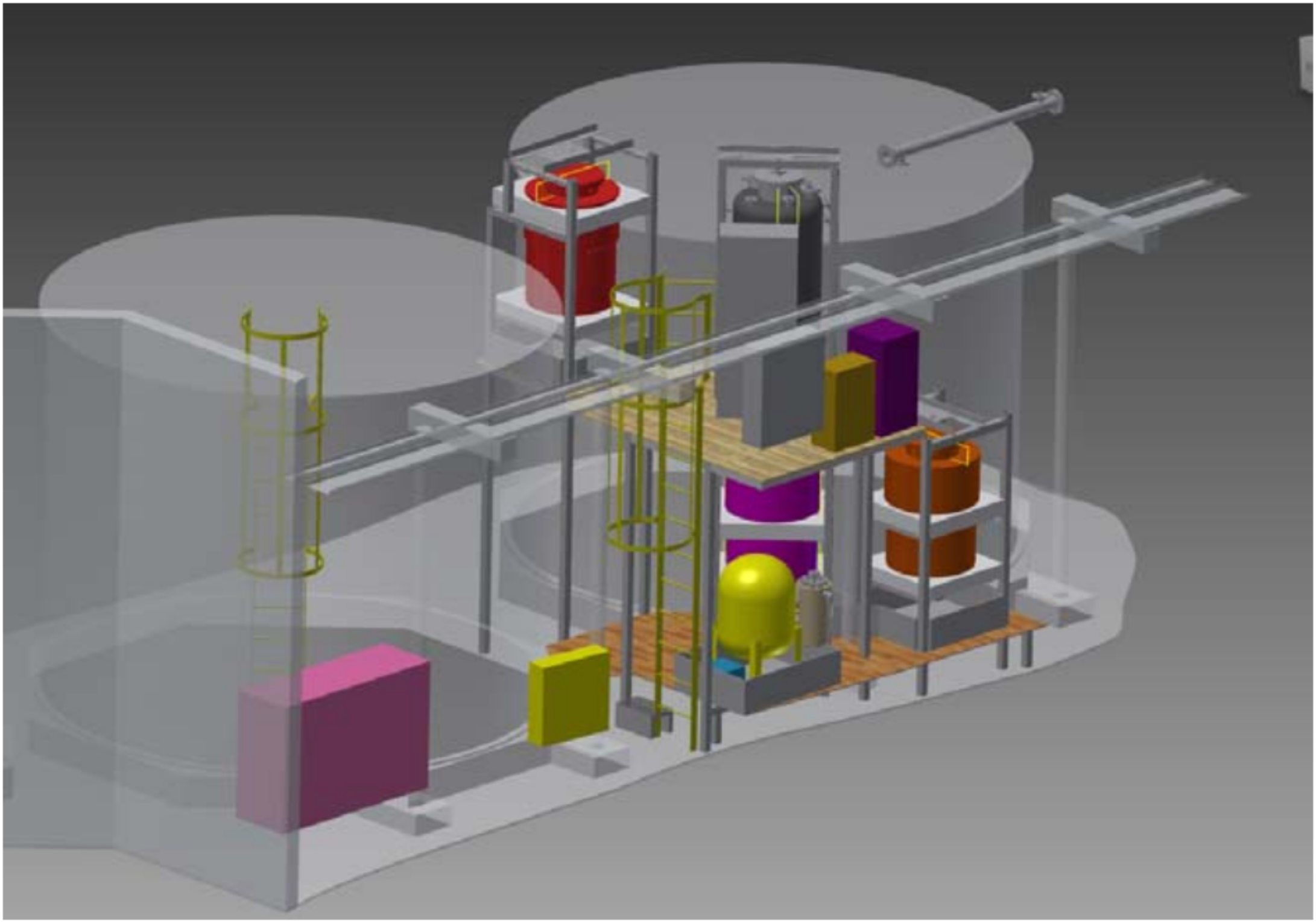}
\caption{\label{fig:TeBDPlant} The current design of the Tellurium Butane Diol Plant}
\end{minipage} 
\end{figure}

Tellurium will be loaded into the scintillator through a Butane-Diol complex (TeBD). A sample of the complex is shown in fig.~\ref {fig:TeBD}.The TeBD production plant will be constructed in the scintillator storage room underground at SNOLAB. A drawing of the plant's design is shown in fig.~\ref{fig:TeBDPlant}. The advantages of TeBD+LAB are a long attenuation length, no inherent optical absorption lines, high light yield, and good $\alpha/\beta$ separation by their decay time.

\section{Backgrounds and Sensitivity}
SNO+'s sensitivity is dependent on the constraint and mitigation of backgrounds, achieved by the purification of scintillator and tellurium. Uranium and thorium present in the scintillator produce $\beta$ decays from $^{214}$Bi and $^{212}$Bi that are close to the Q-value of $^{130}$Te, 2.53~MeV. These appear in coincidence with $^{214}$Po and $^{212}$Po $\alpha$ decays, and can be tagged using timing correlations. $^{208}$Tl gammas (2.614~MeV) from detector materials (ropes, PMTs, AV, H$_2$O) are rejected with a fiducial volume cut at $r=3.5$~m. An asymmetric ROI around the Q-value ([$-0.5\sigma, +1.5\sigma$]), reduces the contamination of $2\nu\beta\beta$ events in the signal region. If the expected reduction levels can be achieved, the background rate in the ROI is $13$ events after one year, dominated by the flat spectrum of $^8$B solar neutrinos.

\begin{figure}[h]
\begin{minipage}{0.4\textwidth}
\includegraphics[width=\textwidth]{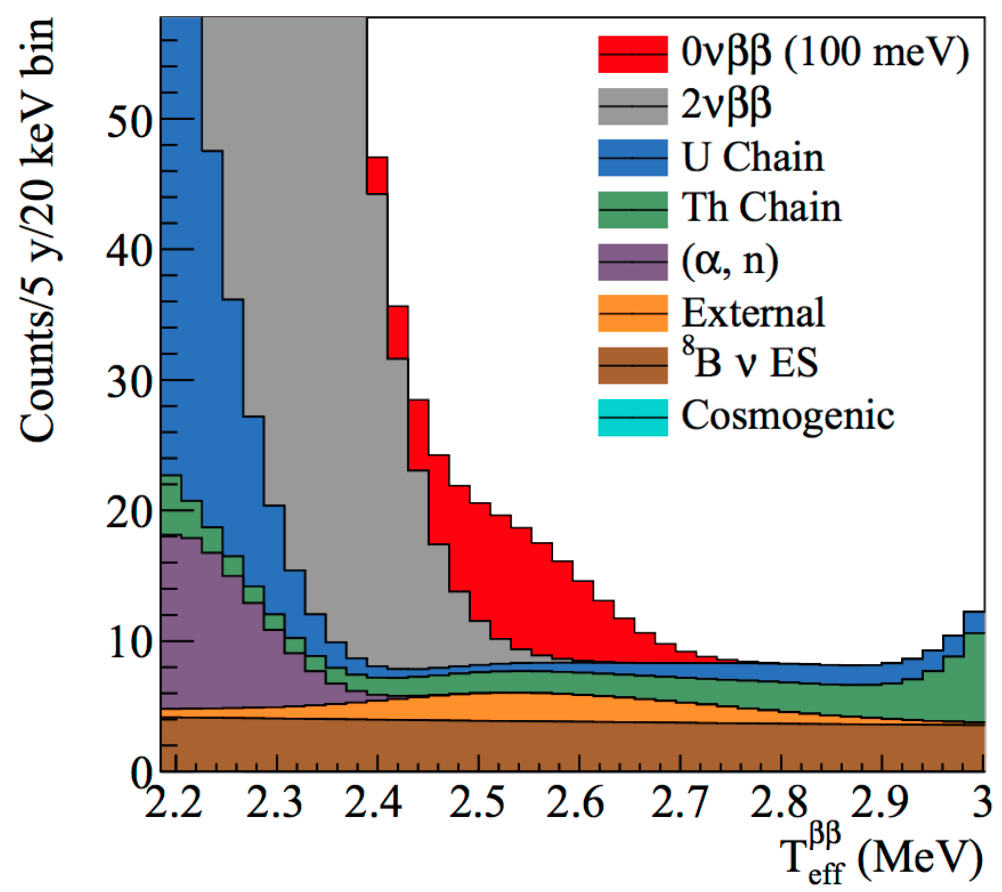}
\caption{\label{fig:0nbbspectrum}Expected energy spectrum for 5 years of SNO+ data for the $^{130}$Te neutrinoless double beta decay search. }
\end{minipage}
\hspace{1pc}%
\begin{minipage}{0.6\textwidth}
\includegraphics[width=\textwidth]{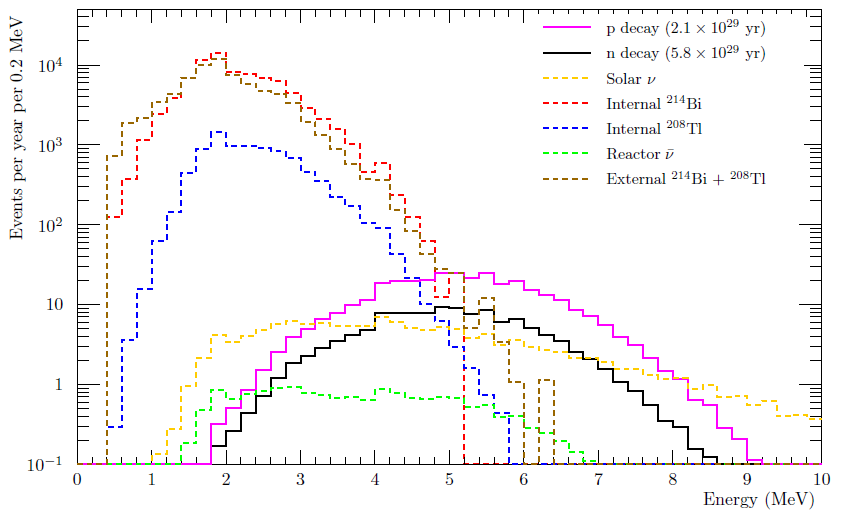}
\caption{\label{fig:ndecspectrum}Expected energy spectrum for 6 months of SNO+ data for the nucleon decay search. }
\end{minipage}\hspace{1pc}%
\end{figure}

Extensive Monte Carlo simulations have been carried out for all backgrounds, in order to estimate their rates and rejection efficiencies. Our baseline assumptions for the study of double beta decay expected sensitivity are a fiducial volume of 20\%, a rejection efficiency of more than 99.99\% for $^{214}$BiPo and 98\% for $^{212}$BiPo, 0.5\% Tellurium loading, and a detected scintillation light yield of 390 PMTs hits/MeV. The expected signal rate was calculated assuming the IBM-2 nuclear matrix element (4.03), gA = 1.269 and G = $3.68\times10^{14}$/yr. The spectrum in fig.~\ref{fig:0nbbspectrum} is shown for five years of running and an assumed Majorana neutrino mass of 100~meV. Based on the above assumptions, we will set a limit on the half life $T^{0\nu}_{1/2} > 2\times10^{26}$~y, at the 90\% CL and the neutrino mass $m_{\beta\beta} \approx 40 - 90$ meV. We are studying higher loading and the deployment of high quantum efficiency PMTs for improved sensitivity.

In our current water phase, SNO+ is sensitive to invisible modes of nucleon decay for both the proton and the neutron, improving the current limits of SNO and KamLAND. Six months of data taking will yield 30 background events in the ROI, shown in figure~\ref{fig:ndecspectrum}. The $90\% C.L.$ limit on neutron decay will be $\tau_{n}=1.2\times10^{30}$~years and on proton decay will be $\tau_{p}=1.4\times10^{30}$~years.

\section{Conclusions and Outlook}
SNO+ is filled with light water and currently taking physics data. In 6 months of running, SNO+ will set world leading limits on invisible nucleon decay. Upgrades to the detector for scintillator-phase running are complete. The scintillator process plant is under commission and the tellurium plant is under construction. Both help SNO+ achieve its desired levels of background purification. The neutrinoless double beta decay phase will begin in late 2018.


\section*{References}
\begin{thebibliography}{9}

\bibitem {review} Andringa S, {\it et al}. 2016 {\it Advances in High Energy Physics} {\bf 2016} Article ID 6194250
\bibitem {scintplant} Ford R. 2015 {\it AIP Conf. Proc.} {\bf 1672} 080003
\bibitem {ropenet} Bialek A, {\it et al}. 2016 {\it NIMA} {\bf 827} 152--160
\bibitem {tellie} Alves R, {\it et al}. 2015 {\it JINST} {\bf 10} P03002
\bibitem {tepurification} Hans S, {\it et al}. 2015 {\it NIMA} {\bf 795} 132--9

\end{thebibliography}
\end{document}